\documentclass[twocolumn,twoside,slac_two]{revtex4}
\usepackage{graphicx}
\usepackage{fancyhdr}
\usepackage{epsf}  % for embedding postscript figures
\usepackage{epsfig}  % for those that use epsfig instead
\usepackage{array}
\usepackage{rotating}
\input babarsym
\RequirePackage{xspace}

\newcommand{\pvec}{{\bf p}}
\newcommand{\acp}{\ensuremath{\calA_{ch}}}
\newcommand{\fL}{\ensuremath{f_L}}
\newcommand{\fzero}{\ensuremath{f_0}}

\newcommand{\calB}{\ensuremath{{\cal B}}}

\newcommand{\timesix}{\ensuremath{\times10^{-6}}}
\newcommand{\pbar}{\kern 0.2em\overline{\kern -0.2em p}{}\xspace}

\newcommand{\DE}{\ensuremath{\Delta E}}

\newcommand\etal{{\it et al.}}
\newcommand{\half}{\ensuremath{{1\over2}}}

\newcommand{\bfig}{\begin{figure}[htbpc!]}
\newcommand{\efig}{\end{figure}}
\newcommand\bef{\begin{figure}}
\newcommand\edf{\end{figure}}
\newcommand\dbline{\noalign{\vskip 0.10truecm\hrule}\noalign{\vskip 2pt}\noalign{\hrule\vskip 0.10truecm}}

\newcommand\beq{\begin{equation}}
\newcommand\eeq{\end{equation}}
\newcommand\bear{\begin{array}}
\newcommand\enar{\end{array}}
\newcommand\beqa{\begin{eqnarray}}
\newcommand\eeqa{\end{eqnarray}}
\newcommand\ben{\begin{enumerate}}
\newcommand\een{\end{enumerate}}

\newcommand{\UfourS}{\ensuremath{\Upsilon(4S)}}
\newcommand{\aone}{\ensuremath{a_1}}

\newcommand{\etaptoepp}{\ensuremath{\etapr\ra\eta\pip\pim}}

\newcommand{\etaptorg}{\ensuremath{\etapr\ra\rho^0\gamma}}

\newcommand{\Kstp}{\ensuremath{\Kstarp}}
\newcommand{\Kstz}{\ensuremath{\Kstarz}}
   \newcommand{\rhop}{\ensuremath{\rho^+}}
   
   \newcommand{\rhom}{\ensuremath{\rho^-}}
   
   \newcommand{\rhoz}{\ensuremath{\rho^0}}

\newcommand{\fetaKst}{\ensuremath{\eta K^{*}}}
\newcommand{\etaKst}{\ensuremath{\B\ra\fetaKst}}

\newcommand{\fetapK}{\ensuremath{\etapr K}}
\newcommand{\etapK}{\ensuremath{\B\ra\fetapK}}

\newcommand{\fetappip}{\ensuremath{\etapr\pip}}
\newcommand{\etappip}{\ensuremath{\Bp\ra\fetappip}}

\newcommand{\fetapKp}{\ensuremath{\etapr K^+}}
\newcommand{\etapKp}{\ensuremath{\Bp\ra\fetapKp}}

\newcommand{\fetappiz}{\ensuremath{\etapr\piz}\xspace}
\newcommand{\etappiz}{\ensuremath{\Bz\ra\fetappiz}\xspace}

\newcommand{\etapKst}{\ensuremath{\B\ra\etapr K^{*}}\xspace}

\newcommand{\fetapKstp}{\ensuremath{\etapr K^{*+}}}
\newcommand{\etapKstp}{\ensuremath{\Bp\ra\etapr K^{*+}}}

\newcommand{\retapKstp}{\ensuremath{xx^{+xx}_{-xx}\pm xx}}

\newcommand{\uletapKstp}{\ensuremath{xx}}

\newcommand{\setapKstp}{\ensuremath{xx}}
\newcommand{\fetapKstz}{\ensuremath{\etapr K^{*0}}}
\newcommand{\etapKstz}{\ensuremath{\Bz\ra\fetapKstz}}

\newcommand{\retapKstz}{\ensuremath{xx^{+xx}_{-xx}\pm xx}}

\newcommand{\setapKstz}{\ensuremath{xx}}
\newcommand{\fetaprho}{\ensuremath{\etapr\rho}\xspace}

\newcommand{\etaprho}{\ensuremath{\B\ra\fetaprho}\xspace}

\newcommand{\fetaprhop}{\ensuremath{\etapr\rho^+}}

\newcommand{\retaprhop}{\ensuremath{xx^{+xx}_{-xx}\pm xx}}

\newcommand{\uletaprhop}{\ensuremath{xx}}

\newcommand{\setaprhop}{\ensuremath{xx}}
\newcommand{\fetaprhoz}{\ensuremath{\etapr\rho^0}}

\newcommand{\retaprhoz}{\ensuremath{xx^{+xx}_{-xx}\pm xx}}

\newcommand{\uletaprhoz}{\ensuremath{xx}}

\newcommand{\setaprhoz}{\ensuremath{xx}}
\newcommand{\fetapfz}{\ensuremath{\etapr f_0}\xspace}
\newcommand{\etapfz}{\ensuremath{\Bz\ra\fetapfz}\xspace}

\newcommand{\retapfz}{\ensuremath{xx^{+xx}_{-xx}\pm xx}\xspace}

\newcommand{\uletapfz}{\ensuremath{xx}\xspace}

\newcommand{\setapfz}{\ensuremath{xx}\xspace}

\newcommand{\fomegapip}{\ensuremath{\omega\pi^+}}
\newcommand{\omegapip}{\ensuremath{\Bp\ra\fomegapip}}

\newcommand{\romegapip}{\ensuremath{xx \pm xx\pm xx}}

\newcommand{\Aomegapip}{\ensuremath{0.xx\pm 0.yy \pm 0.zz}}

\newcommand{\somegapip}{\ensuremath{xx}}

\newcommand{\fomegaKp}{\ensuremath{\omega K^+}}
\newcommand{\omegaKp}{\ensuremath{\Bp\ra\fomegaKp}}

\newcommand{\romegaKp}{\ensuremath{xx\pm xx\pm xx}}

\newcommand{\AomegaKp}{\ensuremath{0.xx\pm 0.yy \pm 0.zz}}

\newcommand{\somegaKp}{\ensuremath{xx}}

\newcommand{\fomegaKz}{\ensuremath{\omega K^0}}
\newcommand{\fomegaKs}{\ensuremath{\omega\KS}}

\newcommand{\omegaKs}{\ensuremath{\Bz\ra\fomegaKs}}

\newcommand{\romegaKz}{\ensuremath{5.9^{+1.0}_{-0.9}\pm xx}}

\newcommand{\somegaKz}{\ensuremath{xx}}

\providecommand{\fomegaKstp}{\ensuremath{\omega\Kstp}}

\providecommand{\romegaKstp}{\ensuremath{3.5^{+2.5}_{-2.0}\pm0.7}}

\providecommand{\ulomegaKstp}{\ensuremath{7.4}}

\providecommand{\fomegaKstz}{\ensuremath{\omega\Kstz}}
\providecommand{\omegaKstz}{\ensuremath{\Bz\ra\fomegaKstz}}

\newcommand{\romegaKstz}{\ensuremath{3.4^{+1.8}_{-1.6}\pm0.4}}

\providecommand{\ulomegaKstz}{\ensuremath{6.1}}

\newcommand{\fomegarhop}{\ensuremath{\omega\rho^+}\xspace}
\newcommand{\omegarhop}{\ensuremath{\Bp\ra\fomegarhop}\xspace}

\newcommand{\romegarhop}{\ensuremath{13.5^{+3.8}_{-3.5}\pm1.7}\xspace}

\newcommand{\fomegarhoz}{\ensuremath{\omega\rho^0}}

\newcommand{\romegarhoz}{\ensuremath{0.59^{+1.3}_{-1.1}\pm0.35}}

\newcommand{\ulomegarhoz}{\ensuremath{3.3}\xspace}

\newcommand{\fomegafz}{\ensuremath{\omega f_0}\xspace}

\newcommand{\fomegaomega}{\ensuremath{\omega\omega}\xspace}
\newcommand{\omegaomega}{\ensuremath{\Bz\ra\fomegaomega}\xspace}

\newcommand{\fomegaphi}{\ensuremath{\omega\phi}\xspace}

\renewcommand{\retapKstz}{\ensuremath{3.8\pm1.1\pm0.5}}

\renewcommand{\setapKstz}{\ensuremath{4.5}}
\renewcommand{\retapKstp}{\ensuremath{4.9^{+1.9}_{-1.7}\pm0.8}}
\renewcommand{\uletapKstp}{\ensuremath{7.9}}
\renewcommand{\setapKstp}{\ensuremath{3.6}}
\renewcommand{\retaprhop}{\ensuremath{6.8^{+3.2}_{-2.9}{}^{+3.9}_{-1.3}}}
\renewcommand{\uletaprhop}{\ensuremath{14}}
\renewcommand{\setaprhop}{\ensuremath{2.3}}
\renewcommand{\retaprhoz}{\ensuremath{0.4^{+1.2}_{-0.9}{}^{+1.6}_{-0.6}}}
\renewcommand{\uletaprhoz}{\ensuremath{3.7}}
\renewcommand{\setaprhoz}{\ensuremath{0.3}}
\renewcommand{\retapfz}{\ensuremath{0.1^{+0.6}_{-0.4}{}^{+0.9}_{-0.4}}}
\renewcommand{\uletapfz}{\ensuremath{1.5}}
\renewcommand{\setapfz}{\ensuremath{0.2}}

\renewcommand{\romegaKstz}{\ensuremath{2.4\pm1.1\pm0.7}}
\renewcommand{\romegaKstp}{\ensuremath{0.6^{+1.4+1.1}_{-1.2-0.9}}}
\renewcommand{\romegarhop}{\ensuremath{10.6\pm2.1^{+1.6}_{-1.0}}}
\renewcommand{\romegarhoz}{\ensuremath{-0.6\pm0.7^{+0.8}_{-0.3}}}
\newcommand{\romegafz}{\ensuremath{0.9\pm0.4^{+0.2}_{-0.1}}}
\newcommand{\romegaomega}{\ensuremath{1.8^{+1.3}_{-0.9}\pm0.4}}
\newcommand{\romegaphi}{\ensuremath{0.1\pm0.5\pm0.1}}
\renewcommand{\ulomegaKstz}{\ensuremath{4.2}}
\renewcommand{\ulomegaKstp}{\ensuremath{3.4}}
\renewcommand{\ulomegarhoz}{\ensuremath{1.5}}
\newcommand{\ulomegaomega}{\ensuremath{4.0}}
\newcommand{\ulomegaphi}{\ensuremath{1.2}}
\newcommand{\ulomegafz}{\ensuremath{1.5}}

\renewcommand{\somegapip}{\ensuremath{10.8}}
\renewcommand{\somegaKp}{\ensuremath{13.0}}
\renewcommand{\somegaKz}{\ensuremath{8.6}}
\renewcommand{\romegapip}{\ensuremath{6.1\pm 0.7\pm 0.4}}
\renewcommand{\romegaKp}{\ensuremath{6.1\pm 0.6\pm 0.4}}
\renewcommand{\romegaKz}{\ensuremath{6.2\pm 1.0\pm 0.4}}
\renewcommand{\Aomegapip}{\ensuremath{-0.01\pm0.10\pm0.01}}
\renewcommand{\AomegaKp}{\ensuremath{0.05\pm0.09\pm0.01}}

\begin{document}

\title{\boldmath{Review of New Rare Hadronic $B$-decay Results}}

\author{James G. Smith}
\affiliation{University of Colorado, Boulder, CO 80309-0390}

\begin{abstract}
We present one new result from Belle and many new results from \babar\
for rare hadronic $B$ decays.  These include measurements of decays 
involving baryons, a Dalitz plot analysis of the three-charged-kaon system,
many new results for $B$ decays to $\etapr X$ and $\omega X$, and a
limit for the decay $B\to a_1\rho$.  Measurements of the vector-vector decays 
$B\to\rho\Kstar$ and $B\to\omega\Kstar$ are helping to understand the value of 
the longitudinal polarization fraction for these $B\to VV$ decays.
\end{abstract}

\maketitle

\thispagestyle{fancy}

\section{Introduction}
In this paper we cover dozens of new measurements of branching fractions,
charge asymmetries and longitudinal polarization of rare hadronic decays
of $B$ mesons.  We will summarize the new results and compare with
previous results and theoretical expectations.

\section{\boldmath{$\Bzb\to D^{*+}\omega\pi^-$}}
\begin{figure}[!htb]
\centerline{\includegraphics[angle=0,scale=0.4]{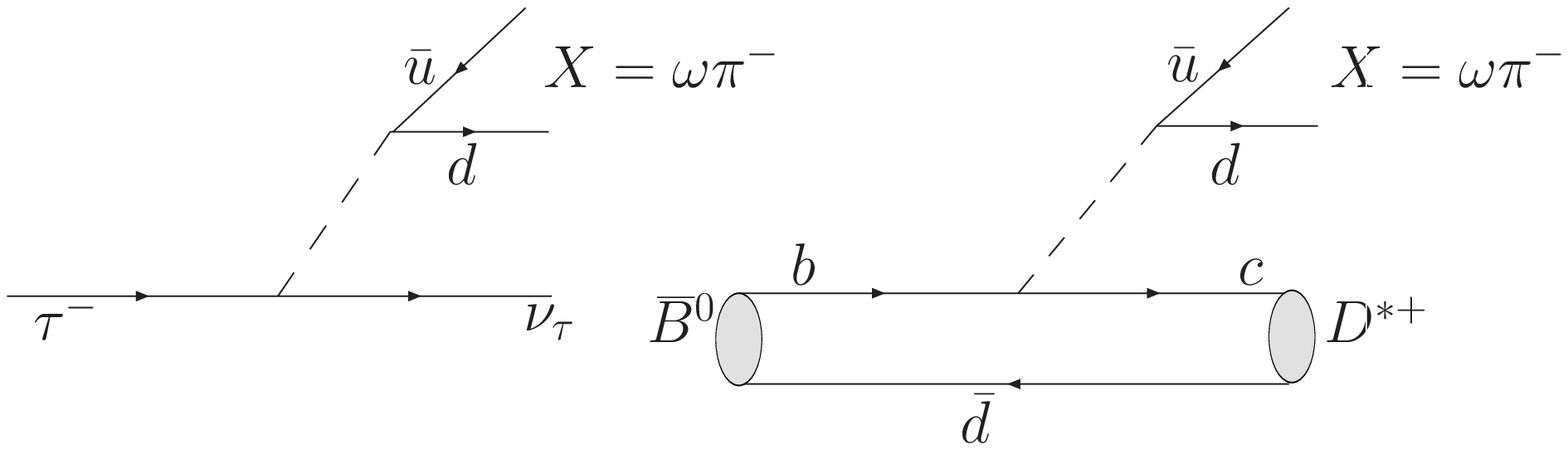}}
\vspace{-.05cm}
\caption{\label{dstompi_feyn}
Feynman diagrams for $\tau^-\to\omega\pi^-\nu_\tau$ and $\Bzb\to
D^{*+}\omega\pi^-$.}
\end{figure}

A new measurement from \babar\ of the decay $\Bzb\to D^{*+}\omega\pi^-$
\cite{Dstomegapi} is a test of factorization since the amplitude is related 
to the amplitude for the decay $\tau^-\to\omega\pi^-\nu_\tau$ \cite{liluwi} 
(see Fig. \ref{dstompi_feyn}).
Charge-conjugate decay modes are implied throughout this paper unless
explicitly stated otherwise.  

\begin{figure}[!htb]
\centerline{\includegraphics[angle=0,scale=0.4]{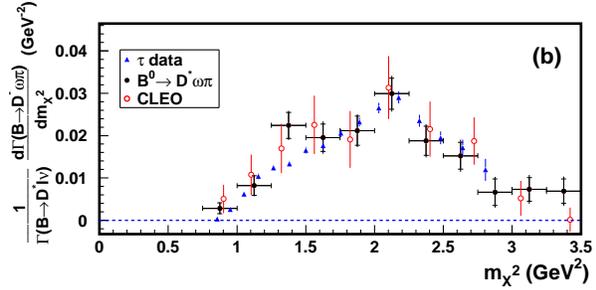}}
\vspace{-.2cm}
 \caption{\label{dstompi_masssq}
Data for the $\Bzb\to D^{*+}\omega\pi^-$ differential distribution of 
$\omega\pi$ mass squared, normalized to the
semileptonic width $\Gamma(B \to D^* \ell \nu)$.  The total
error bars include the $m_X^2$-dependent systematic uncertainties but
not a common 11.3\% scale systematic uncertainty.  Also shown are the
predictions from CLEO $\tau$ data and the previous CLEO analysis of
this $B$  decay.}
\end{figure}

In Fig. \ref{dstompi_masssq}, we show
the normalized differential distribution for the $\omega\pi$ mass squared.  We
compare with the CLEO $\tau$ data \cite{CLEOtau} and a measurement by
CLEO of the differential $B$ decay spectrum \cite{CLEODstomegapi}.
There is good agreement with the CLEO data and the factorization expectation.

\section{Decays with baryons}

A variety of measurements of $B$ decays to baryonic final states, until
recently mostly by Belle, has challenged theoretical understanding of
these decays.  Issues such as threshold enhancements and angular
correlation are still not completely understood.  There are two new
measurements from \babar\ that will be discussed in the next sections.

\subsection{\boldmath{$\Bzb\to\Lambda_c^+ \pbar$}}

\begin{figure}[!htb]
 \includegraphics[angle=0,scale=0.4]{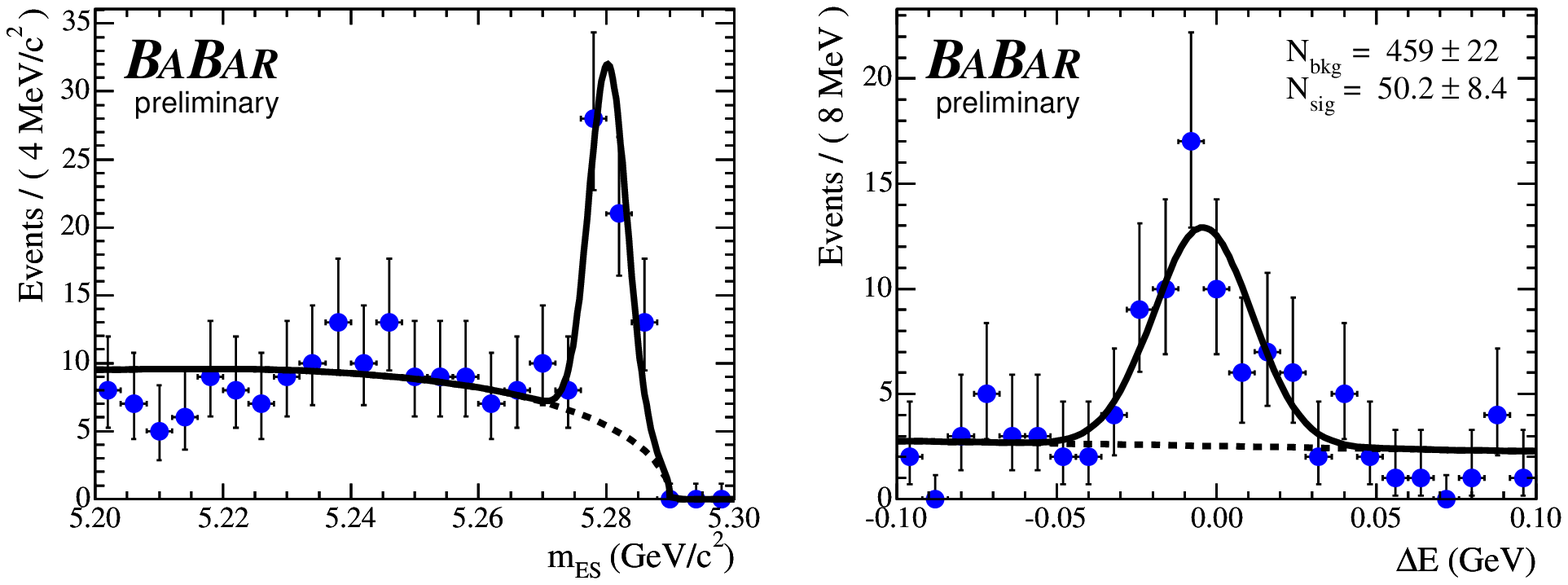}
\vspace{-.2cm}
 \caption{\label{Lambdacpbar}Projection plots of the \mes\ and \DE\
distributions for the $\Bzb\to\Lambda_c^+ \pbar$ analysis.}
%\vspace{-.4cm}
\end{figure}

\babar\ has measured the branching fraction for the decay $\Bzb\to\Lambda_c^+\pbar$.  
They perform a maximum-likelihood (ML) fit to the quantities \mes\
and \DE, where $\mes\equiv\sqrt{(\half s + \pvec_0\cdot\pvec_B)^2/E_0^2 
- \pvec_B^2}$ and the energy difference $\DE \equiv E_B^*-\half\sqrt{s}$, where
$(E_0,\pvec_0)$ and $(E_B,\pvec_B)$ are four-momenta of
the \UfourS\ and the $B$ candidate, respectively, and the asterisk
denotes the \UfourS\ rest frame.  The observed signal is $50.2\pm8.4$
events (see Fig. \ref{Lambdacpbar}), leading to a branching fraction of 
$(2.15\pm0.36\pm0.13\pm0.56)\times 10^{-5}$, where the uncertainties are,
respectively, statistical, systematic and in the $\Lambda_c^+\to
pK^-\pip$ branching fraction.  This result is in good agreement
with the published Belle measurement \cite{BelleLambdacpbar} and theoretical
expectations \cite{CY}.

\subsection{\boldmath{$\Bzb\to\Lambda\pbar\pip$}}

\begin{figure}[!htb]
 \includegraphics[angle=0,scale=0.37]{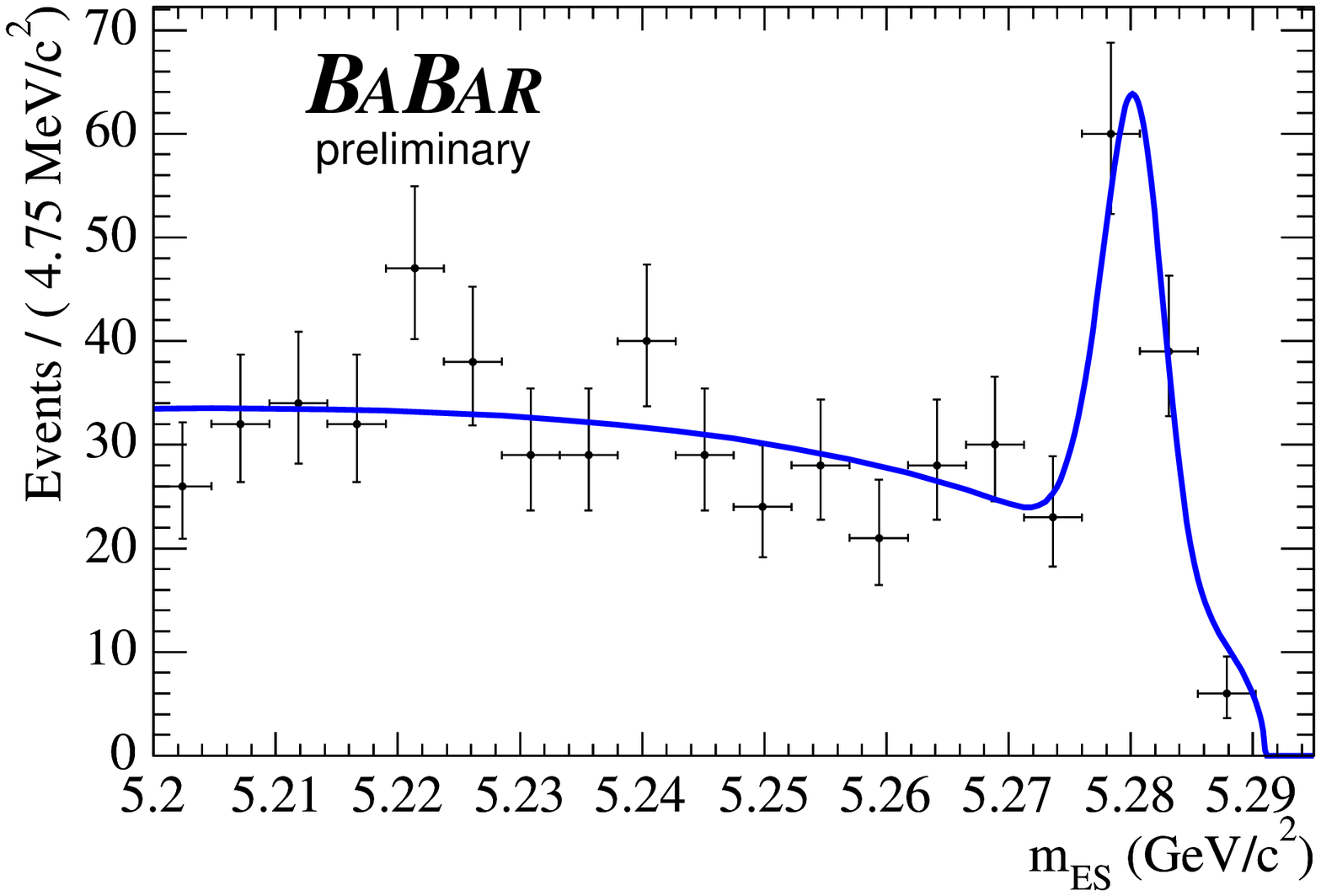}
 \includegraphics[angle=0,scale=0.37]{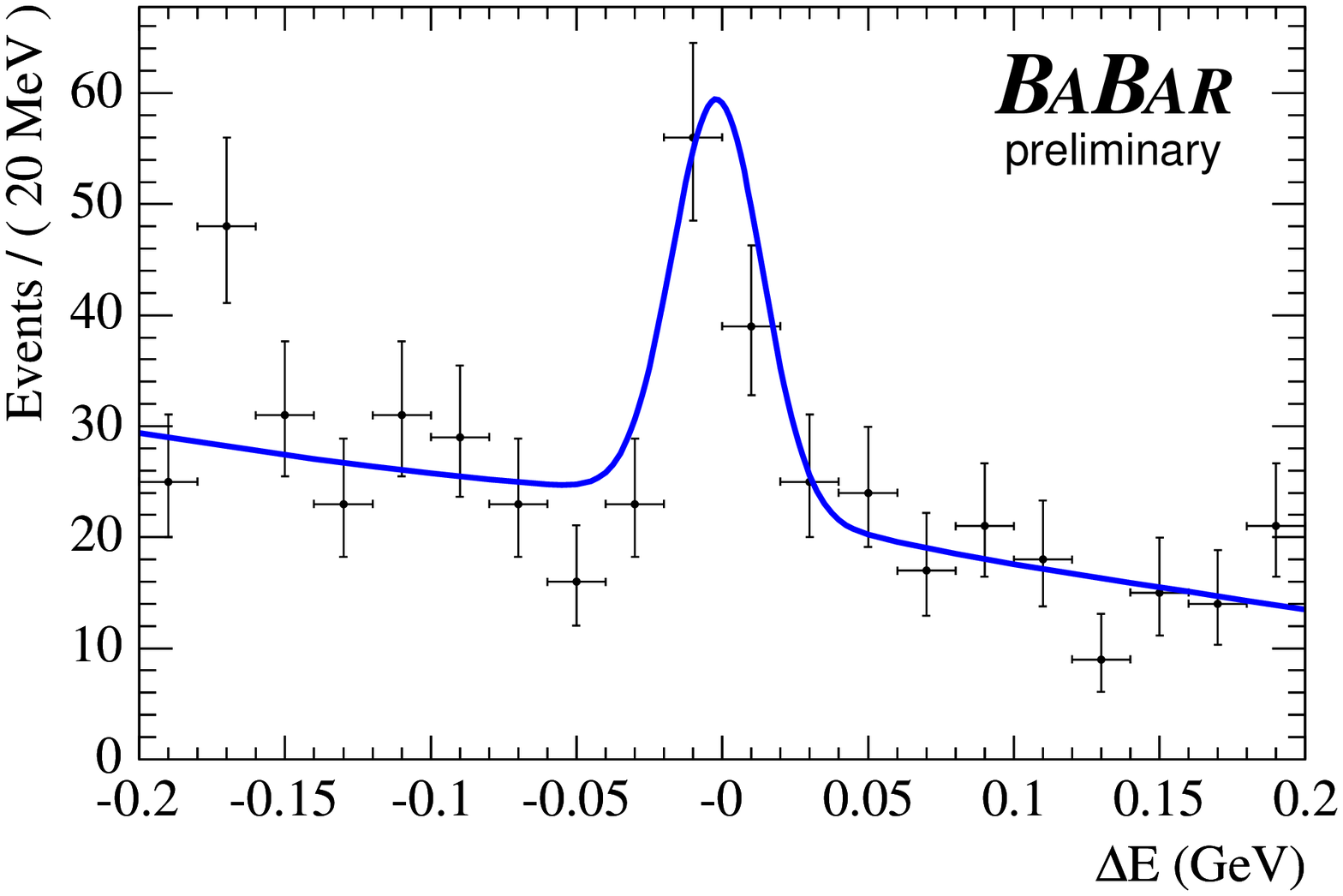}
\vspace{-.2cm}
 \caption{\label{Lambdappi_mesDE}Projection plots of the \mes\ and \DE\
distributions for the $\Bzb\to\Lambda\pbar\pip$ analysis.}
%\vspace{-.4cm}
\end{figure}

\begin{figure}[!htb]
\centerline{\includegraphics[angle=0,scale=0.4]{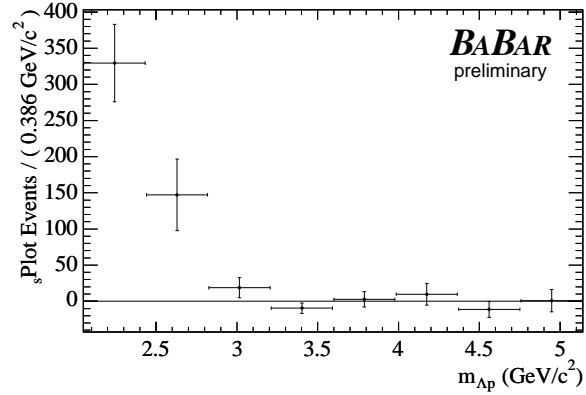}}
\vspace{-.2cm}
 \caption{\label{Lambdappi_mlamp}Plot of the $\Lambda\pbar$ invariant
mass showing the enhancement at threshold.}
%\vspace{-.4cm}
\end{figure}

Another new \babar\ result is for the decay $\Bzb\to\Lambda\pbar\pip$.  
Again a ML fit to \mes\ and \DE\ is used to extract a signal of $\sim74$
events leading to a branching fraction $(3.30\pm0.53\pm0.31)\timesix$.
Figure \ref{Lambdappi_mesDE} shows the projections of the signal onto
the \mes\ and \DE\ axes.  Figure \ref{Lambdappi_mlamp} shows that there 
is an enhancement near threshold in the $\Lambda\pbar$ mass.  This
feature is likely important in understanding the relatively large
branching fraction for this decay \cite{HouSoni,ChuaHou}.  This measurement
is also in good agreement with the published Belle result \cite{BelleLambdapbarpi}.

\section{\boldmath{$B^+\to K^+K^-K^+$}}

The new \babar\ measurement of the $B^+\to K^+K^-K^+$ Dalitz plot has now 
been submitted for publication \cite{KKK} so I will not go into the details 
here.  The analysis includes contributions from the final states $\phi K$,
$f0(980) K$, $f0(1710) K$, and $\chi_{c0}K$ and non-resonant as well as
a channel denoted $X_0(1550)K$, previously seen by Belle \cite{BelleKKK},
but not yet understood in terms of known resonances.
The branching fraction for the three-charged-kaon state is measured to be
$(35.2 \pm 0.9 \pm 1.6)\timesix$, somewhat larger than, but in agreement
with the Belle result.  The branching fractions for the various
components of the Dalitz plot are measured.  There is general agreement
with Belle except that the width of the $X_0(1550)$ state is nearly
twice as large in the \babar\ analysis; this disagreement is not
understood.  The consequence is that the allocation of the branching fraction 
between $X_0(1550)K$ and the non-resonant component is different between
the two analyses.  Further progress will surely require a better
understanding of the $X_0(1550)$ state.

\section{\boldmath{New results involving $\etapr$ mesons}}
The decay \etapK\ was discovered nearly ten years ago by CLEO \cite{CLEOetapk} 
with a branching fraction much larger than expected.  This is more or less
understood now as a result of a variety of enhancements including the
effect of $\eta/\etapr$ mixing \cite{Lipkin} and those due to the leading 
order in $1/m_b$ \cite{BN}.  More recently in a SCET calculation, this 
enhancement is identified with an additional term involving 
the gluonic content of the \etapr\ \cite{WZ}.  The prediction that
the $\eta K$ channel is suppressed has been confirmed with fairly
precise recent measurements \cite{BABARetak,BelleetaK}.  The situation
for the other related decays is less clear.  The pattern for the decays
involving $\Kstar$ mesons was originally thought to be reversed due to a
sign flip involving a parity argument \cite{Lipkin}. QCD factorization 
calculations \cite{BN} showed that the situation is more complicated with no
sign flip for the usual QCD penguin amplitudes but a sign flip present
for the $1/m_b$-suppressed non-$V$$-$$A$ amplitudes.  The latter subtleties
have not been widely recognized previously (including by the author in his
verbal presentation in Vancouver).  The result is that the decay
\etaKst\ is clearly enhanced but the suppression of the decay \etapKst\
involves the interplay between the two (opposite-sign) amplitudes
mentioned above as well as possible contributions from flavor-singlet
amplitudes.  The fact that this decay being small is related to the
abnormally large $1/m_b$-suppressed amplitudes was one of the less
appreciated aspects of the QCD factorization calculations \cite{BN}.
The non-strange decays involving a $\pi$ or $\rho$ meson are also interesting.
New measurements of all of several of these decays are discussed in the
following sections.

\subsection{\boldmath{\etapKst\ and \etaprho}}

\begin{figure}[!htb]
 \includegraphics[angle=0,scale=0.2]{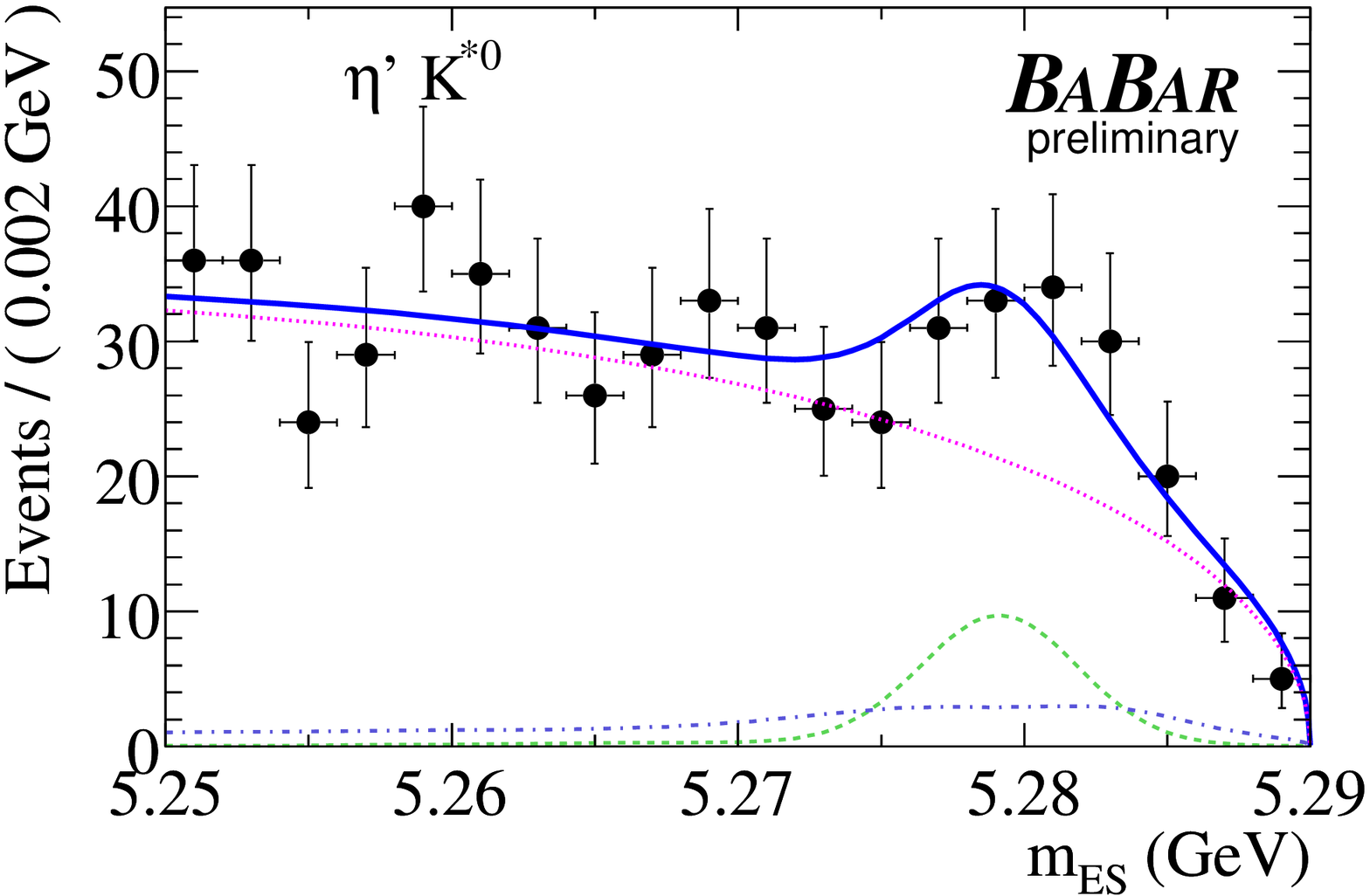}
 \includegraphics[angle=0,scale=0.2]{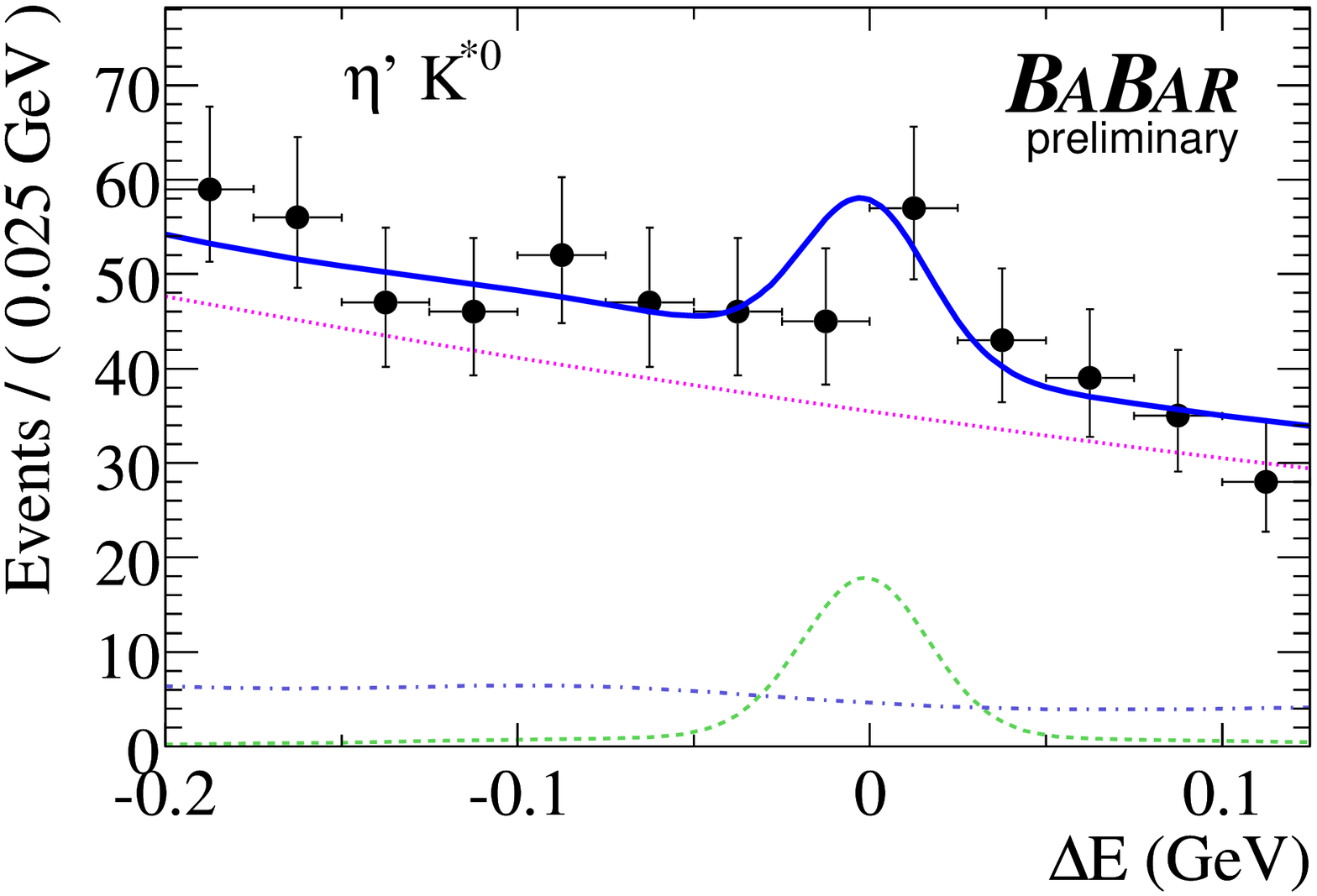}
\vspace{-.2cm}
 \caption{\label{etapKstz_proj}Projection plots of the \mes\ and \DE\
distributions for the $\etapKstz$ analysis.}
%\vspace{-.4cm}
\end{figure}

\babar\ has new results for the decays \etapKst, \etaprho, and \etapfz,
where the latter is measured since it shares a common $\pip\pim$ final
state with \rhoz.  The \etapr\ mesons are reconstructed from the
\etaptoepp\ and \etaptorg\ decay modes and \Kstar\ mesons are reconstructed 
via $\Kstarz\to K^+\pi^-$, $\Kstarp\to K^+\pi^0$ and $\Kstarp\to\Kz\pi^+$.
ML fits are performed, with the variables \mes, \DE,
resonant masses ($\rho$ or \Kstar), the $\rho$ or \Kstar\ helicity angle,
and a Fisher discriminant to distinguish signal from \qqbar\ background
primarily by event shape.  The results for these fits are shown in
Table \ref{tab:comparison}.  The decay \etapKstz\ is observed with a
significance of 4.5 standard deviations ($\sigma$); the \mes\ and \DE\
projection plots for this mode are shown in Fig.~\ref{etapKstz_proj}.  
There is evidence
for \etapKstp\ at the 2.6$\sigma$ level.  However the branching
fractions are small so there clearly is substantial suppression relative
to the \etaKst\ decays which have a branching fraction $\sim20\timesix$
\cite{HFAG}.  It seems that the $1/m_b$ terms must indeed be large for the
suppression to be this large.

\begin{table}[htb]
\caption{Comparison of new \babar\ results with previous results.
Branching fractions (\calB\ in units of $10^{-6}$), significance $S$ and
90\% C.L. upper limits (U.L.) where signal is not significant.}
\label{tab:comparison}
\vspace{-.1cm}
\begin{center}
\setlength{\extrarowheight}{4pt}   % make table rows clearer.
\begin{tabular}{lcccll} 
\dbline
Mode & \multicolumn{2}{c}{Previous results}&\multicolumn{3}{c}{New \babar\ results}\\ 
           &\babar&Belle&\calB &~~$S$~~& \calB\ U.L. \\
\hline
\fetapKstz  & $ <7.6$&$<20$&\retapKstz &$\setapKstz\,\sigma$ &\\
\fetapKstp  & $ <14 $&$<90$&\retapKstp &$\setapKstp\,\sigma$ &$<\uletapKstp$\\
\fetaprhoz  & $ <4.3$&$<14$&\retaprhoz &$\setaprhoz\,\sigma$ &$<\uletaprhoz$\\
\fetaprhop  & $ <22 $& $-$ &\retaprhop &$\setaprhop\,\sigma$ &$<\uletaprhop$\\
\fetapfz    &  $-$   & $-$ & \retapfz  &$\setapfz\,\sigma$   &$<\uletapfz$ \\
\hline
\end{tabular}
\end{center}
\end{table}

\subsection{\boldmath{\etappip\ and \etappiz}}
\begin{figure}[!htb]
\hspace{-0.3cm}
\epsfxsize 1.6 truein \epsfbox{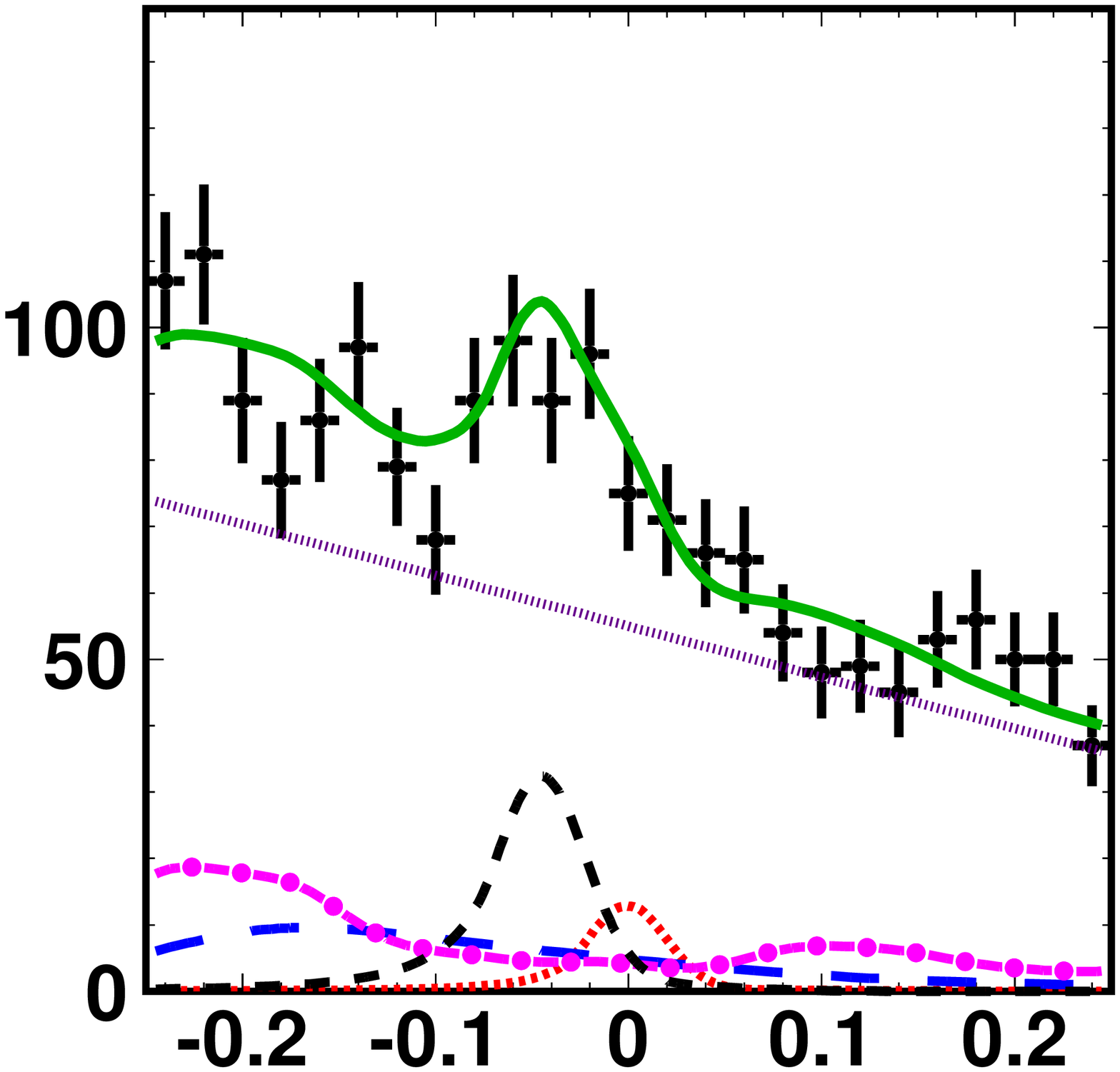}
\epsfxsize 1.6 truein \epsfbox{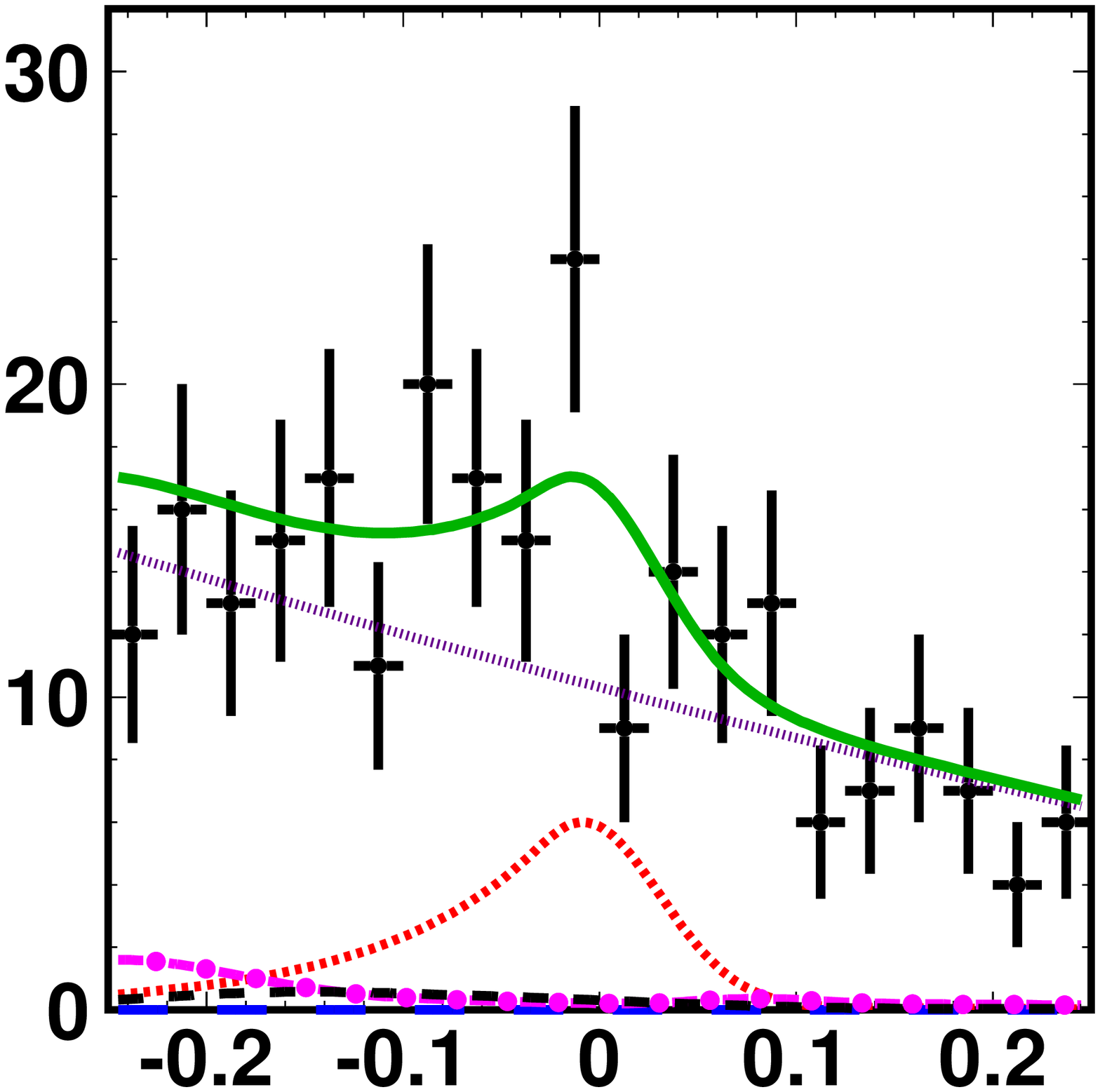}
\put(-65.,-5.0){{\sf\shortstack[c]{\DE\ (GeV)}}}
\put(-185.,-5.0){{\sf\shortstack[c]{\DE\ (GeV)}}}
\put(-240.,40.){{\sf\shortstack[c]{\rotatebox{90}{Events/20 MeV}}}}
\put(-120.,40.){{\sf\shortstack[c]{\rotatebox{90}{Events/25 MeV}}}}
\put(-140.,95.){{\sf\shortstack[c]{(a)}}}
\put(-25.,95.){{\sf\shortstack[c]{(b)}}}
\\
\hspace*{-0.3cm}
\epsfxsize 1.6 truein \epsfbox{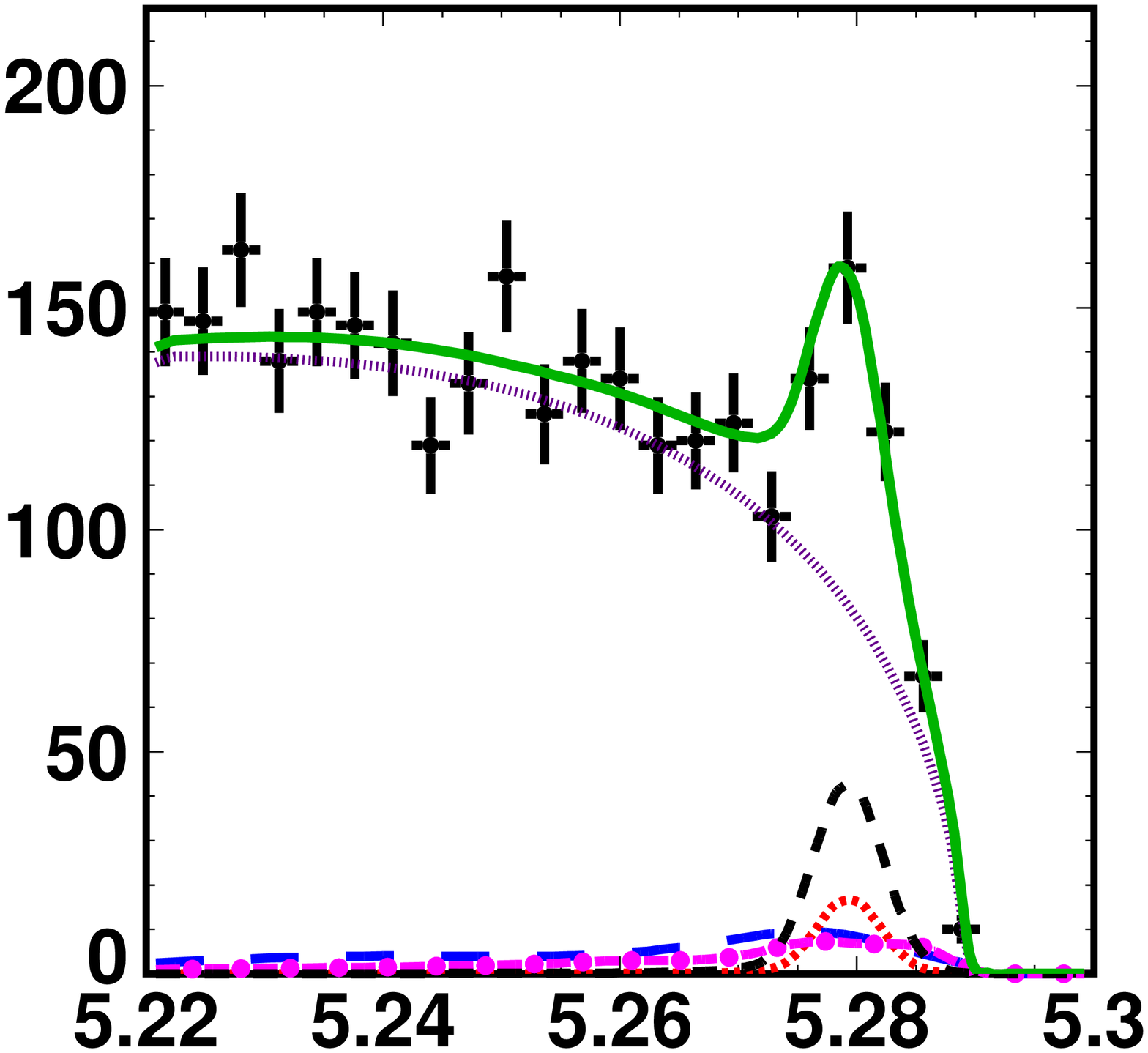}
\epsfxsize 1.6 truein \epsfbox{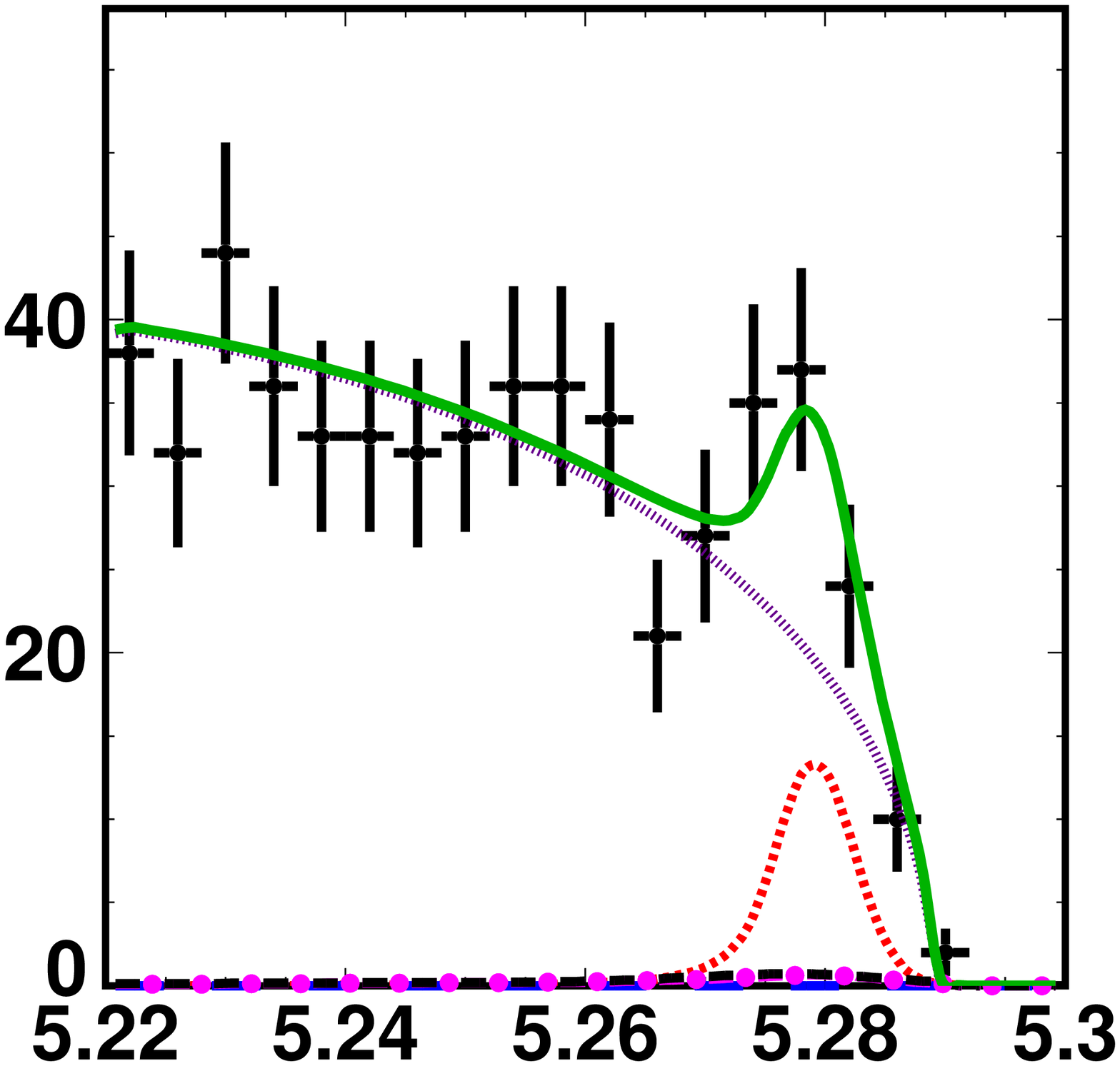}
\put(-65.0,-5.0){{\sf\shortstack[c]{\mes\ (GeV )}}}
\put(-185.,-5.0){{\sf\shortstack[c]{\mes\ (GeV )}}}
\put(-240.,45.){{\sf\shortstack[c]{\rotatebox{90}{Events/4 MeV}}}}
\put(-120.,45.){{\sf\shortstack[c]{\rotatebox{90}{Events/5 MeV}}}}
\put(-140.,95.){{\sf\shortstack[c]{(c)}}}
\put(-25.,95.){{\sf\shortstack[c]{(d)}}}
\caption{\label{Belle_etapi}
Distributions of \DE\ for the Belle (a) \etappip\ and (b) \etappiz\
analyses.  Also \mes\ for (c) \etappip\ and (d) \etappiz.  The signal is
shown as the red (dotted) curve and, for \etappip, the crossfeed from
\etapKp\ is shown as the black dashed curve.}
\end{figure}

Belle has updated the branching fraction
measurements for \etapK\ and \etappip\ and now also measure
\etappiz\ \cite{BELLEetappi}.  They perform ML fits to \mes\ (called $m_{bc}$
by Belle but it is the same quantity) and \DE.  The distributions of \DE\ and 
\mes\ are shown in Fig.~\ref{Belle_etapi}.  They find branching fractions for 
\etappip\ of $(1.8^{+0.8}_{-0.7}\pm0.1)\timesix$ and for \etappiz\
$(2.8\pm1.0\pm0.3)\timesix$.  The significance for the signals is 3.2
and 3.1$\sigma$, respectively.

The new Belle branching fraction for \etappip\ is somewhat smaller than, but
consistent with, the published observation from \babar, 
$(4.0\pm0.8\pm0.4)\timesix$ \cite{BABARetak}.  The situation for
\etappiz\ is reversed -- the new Belle measurement is somewhat larger
than the recently published measurement from \babar,
$(0.8^{+0.8}_{-0.6}\pm0.1)\timesix$ ($<2.1\timesix$) \cite{etapi0}.
While the world-average branching fraction for \etappiz\ is not yet
significant, it seems that this channel too may be close to observation at 
about the expected rate.  This is interesting since the penguin for this mode
is expected to be small and the color-suppressed tree amplitude is also
expected to be strongly suppressed \cite{CGRS}.

\section{Charmless Vector-Vector decays}
The decays of a $B$ meson to pairs of vector particles are interesting
for a variety of reasons.  Early interest focused on issues such as
\CP-violating observables.  In the last few years the focus has been on 
the longitudinal polarization fraction \fL.  This is naively expected
to be close to 1.0 since spin-flip arguments indicate that the transverse
polarization is of the order $m_V^2/m_B^2\sim0.04$.  The measurement of 
\fL\ for the $B\to\phi\Kstar$ decays is now known to be near 0.5 with
an uncertainty of about 0.04 \cite{BABARphikst,Bellephikst}.  This
appears to be unique to decays dominated by penguins, though the exact
mechanism is still not understood despite dozens of theory papers
offering Standard Model \cite{VVBSMrefs} or non-SM explanations~\cite{nSMetc}.

In the following sections we describe a series of measurements from
\babar\ for several of these vector-vector decays.  In addition see the 
\babar\ measurements of branching fraction and polarization for the 
$\Bp\to\rhop\rhoz$ decay are described in the talk by Christos
Touramanis at this conference.

\subsection{\boldmath{Vector-vector decays involving $\omega$ mesons}}

\begin{table*}[!bt]
\caption{\label{tab:omegarho_results}
Quantities measured in the \babar\ $\omega X$ analysis.
Measured branching fraction \calB, significance $S$ (with systematic 
uncertainties included), 90\% C.L. upper limit, measured
or assumed longitudinal polarization, and charge asymmetry \acp.
}
%\vspace{-5mm}
\begin{tabular}{lccccc}
\dbline
Mode      & $\calB\ (10^{-6})$  & $S~(\sigma)$ & \calB\ U.L.$(10^{-6})$ &  ~~$f_L$  & \acp \\
\dbline
\fomegaKstz     & \romegaKstz &2.4& \ulomegaKstz & $0.71^{+0.27}_{-0.24}$ & --- \\
\fomegaKstp     & \romegaKstp &0.4& \ulomegaKstp &0.7 fixed     & --- \\
\fomegarhoz     & \romegarhoz &0.6& \ulomegarhoz &0.9 fixed      & --- \\
\fomegafz       & \romegafz   &2.8& \ulomegafz& ---       & --- \\
\fomegarhop     & \romegarhop &5.7& $-$ &$0.82\pm0.11\pm0.02$&$0.04\pm0.18\pm0.02$ \\
\fomegaomega    & \romegaomega&2.1& \ulomegaomega &0.79$\pm0.34$ & --- \\
\fomegaphi      & \romegaphi  &0.3& \ulomegaphi & 0.88 fixed & --- \\
\dbline
\end{tabular}
\end{table*}

\babar\ has recently submitted for publication an analysis of $B\to VV$
decays where one of the vector mesons is an $\omega$ \cite{omegaV}.  
Unbinned ML fits are performed, with the following variables in the fit: \mes; 
\DE; resonant masses ($\omega$, \Kstar\ or $\rho$); the $\omega$, $\phi$, 
$\rho$ or \Kstar\ helicity angle; and a Fisher discriminant similar to
that used for the \etapKst\ analysis.  The
results of this analysis are summarized in Table \ref{tab:omegarho_results}.
Again the \fzero\ channel is included since it shares a common $\pip\pim$ 
final state with \rhoz.  The only channel with a significant yield is
\omegarhop.  For this case, the longitudinal polarization
fraction and charge asymmetry \acp\ are also measured.  For the 
\omegaKstz\ and \omegaomega\ decays, where there are signal yields of
about 50 events, \fL\ is left free in the fit (though since this is not
considered a measurement of \fL, no systematic error is given).  For the other 
channels, \fL\ is fixed to the approximate expected value and varied by 0.3 to
obtaining systematic errors.  Belle has not yet reported searches for
these decays.

If the $\omega\Kstar$ channels were dominated by penguin diagrams, the
branching fraction would be expected to be one-half of the branching
fraction of $\Bp\to\rhop\Kstarz$ (see next section) or $\sim5\timesix$.
This seems unlikely given the measurements shown in Table
\ref{tab:omegarho_results}.  This suggests that the (Cabibbo-suppressed)
tree amplitude for the $\omega\Kstar$ decays may not be negligible.
This would indicate the possibility for measuring a large value of \acp\
once these decays are observed.

\subsection{\boldmath{$B\to\rho\Kstar$}}
The various charge states of $B\to\rho\Kstar$ are interesting since some
are known to have significant branching fractions and \fL\ can be
measured to compare with $\phi\Kstar$.  

\subsubsection{\boldmath{$\Bp\to\rhop\Kstarz$}}
The decay $\Bp\to\rhop\Kstarz$ is particularly interesting since it is
thought to be a pure penguin (there is no tree diagram for this decay).
In addition to the interest in the polarization, a recent paper
\cite{BGRS} has suggested that the branching fraction and \fL\ from this
decay can be use to limit the penguin uncertainty for the measurement of
the CKM angle $\alpha$ in the decay $\Bz\to\rhop\rhom$.  

\begin{figure}[!htb]
 \includegraphics[angle=0,scale=0.42]{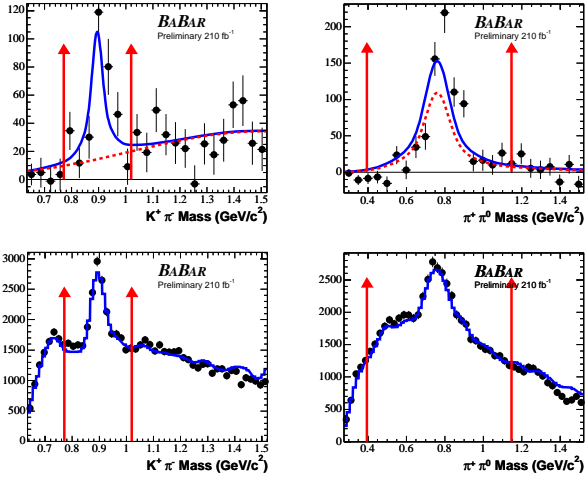}
\vspace{-.4cm}
 \caption{\label{Kstzrhop_widemass}For the $\Bp\to\rhop\Kstarz$ analysis, 
``sPlots" of the $K^+\pim$ mass (left) and $\pip\piz$ mass (right) for signal 
(top) and \qqbar\ background (bottom).  The blue solid curves represent the
full signal or background components and the red dashed curve indicates the 
contribution from S-wave $K\pi$.  The plot range is for the wide fit
region, while arrows indicate the nominal fit range.}
\end{figure}

\begin{figure}[!htb]
 \includegraphics[angle=0,scale=0.42]{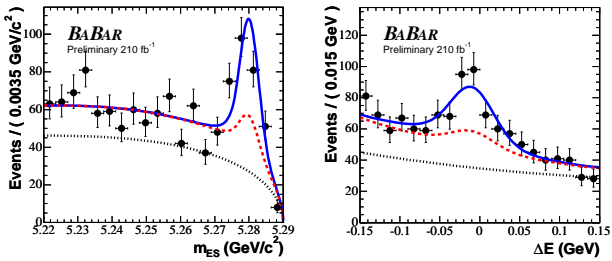}
\vspace{-.5cm}
 \caption{\label{Kstzrhop_proj}Projection plots of the \mes\ and \DE\
distributions for the $\Bp\to\rhop\Kstarz$ analysis.  The black dotted lines
show the \qqbar\ background component, the red dashed lines indicate the
full background component including $K\pi$ S-wave, and the blue solid line
is for the full fit with the signal component.}
\end{figure}

\babar\ has a new measurement for this decay, using a ML analysis with 
inputs \mes, \DE, $\rho$ and \Kstar\ masses, the $\rho$ and \Kstar\ 
helicity angles, and a neural-net variable analogous to the Fisher
discriminant event-shape variable used in other \babar\ analyses.
The $K^+\pim$ decay used to reconstruct the $K^*(892)$ also has peaking
at higher mass due to a combination of $\Kstar_0$(1430) and nearby
non-resonant S-wave signal.  A $K\pi$ mass range extending to 1.5 GeV is used 
to determine the amount of S-wave signal, while a more typical narrow mass 
range is used for the main $K^*(892)$ analysis.  These regions are shown
in the ``sPlots" \cite{sPlot} of Fig.~\ref{Kstzrhop_widemass} where the plot
indicates the wide ranges used and the arrows indicate the narrow ranges.
The signal plot in the top left shows that the S-wave $K\pi$ signal is
substantially larger than the $K^*(892)$ signal.  There is no evidence
for contributions other than \rhop\ in the $\pip\piz$ invariant mass.
In Fig.~\ref{Kstzrhop_proj} we show the $\Bp\to\rhop\Kstarz$ signal of 
$\sim$210 events with the \mes\ and \DE\ projection plots for the 
nominal mass region.  The measured branching fraction is 
$(10.0\pm1.7\pm2.4)\timesix$, $\fL=0.53\pm0.10\pm0.06$, and 
$\acp=-0.01\pm0.15\pm0.01$.  The first two are in good agreement with the
published Belle measurement \cite{BelleKstrho} (Belle has not yet
measured \acp).  The value of \fL\ is in good agreement with the value
for $\phi\Kstar$ as expected for pure penguin decays.

\subsubsection{$\Bp\to\rhoz\Kstarp$}

The decay $\Bp\to\rhoz\Kstarp$ is less clear theoretically because there
is a (Cabibbo-suppressed) tree diagram which contributes in addition to
the penguin present for all $B\to\rho\Kstar$ decays.  It is more
difficult experimentally since the branching fraction is smaller (as for
$B\to\omega\Kstar$, it would be suppressed by a factor of two if penguin 
amplitudes were the only ones contributing).  

\begin{figure}[!tb]
 \includegraphics[angle=0,scale=0.2]{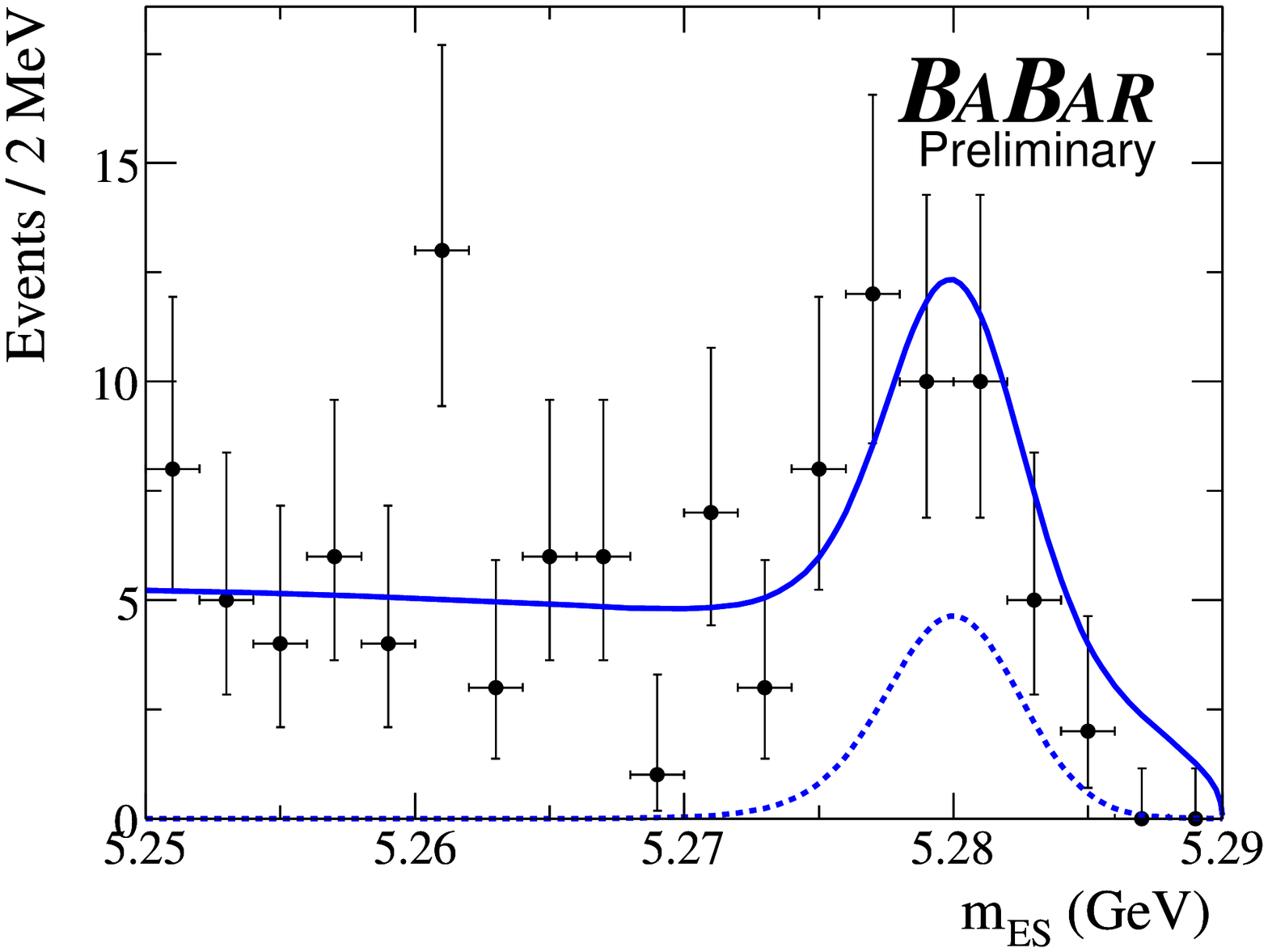}
 \includegraphics[angle=0,scale=0.2]{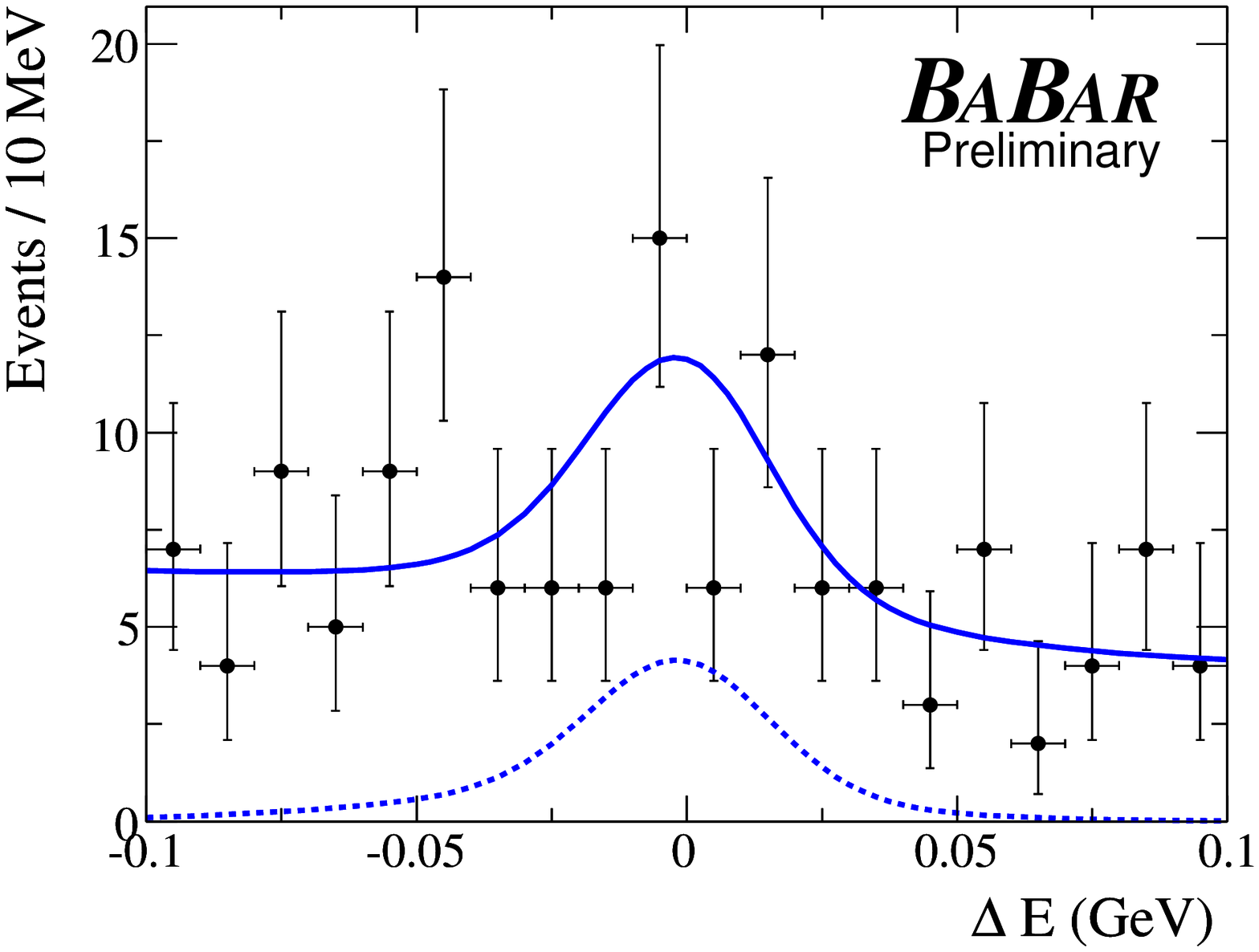}\\
 \includegraphics[angle=0,scale=0.2]{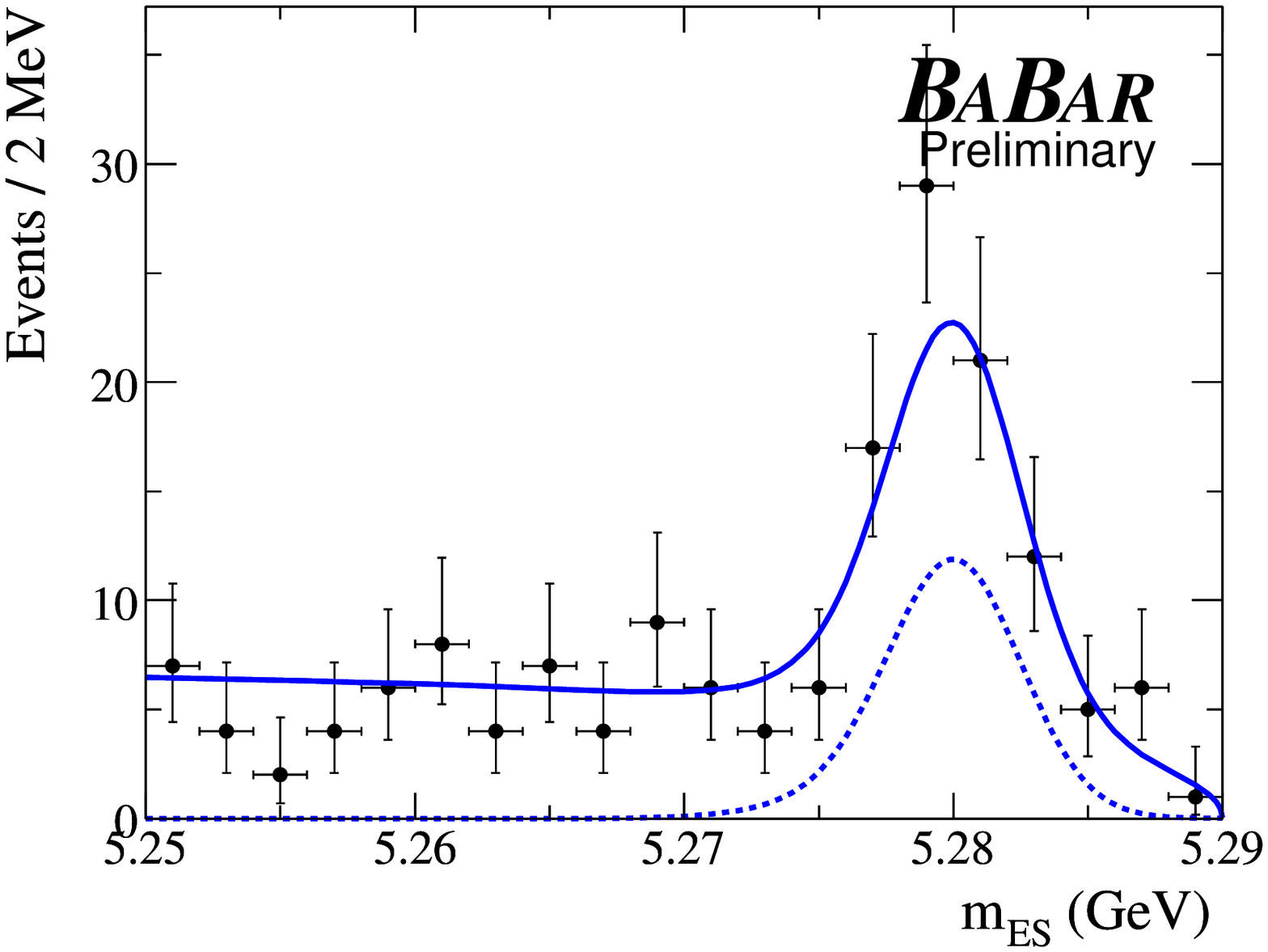}
 \includegraphics[angle=0,scale=0.2]{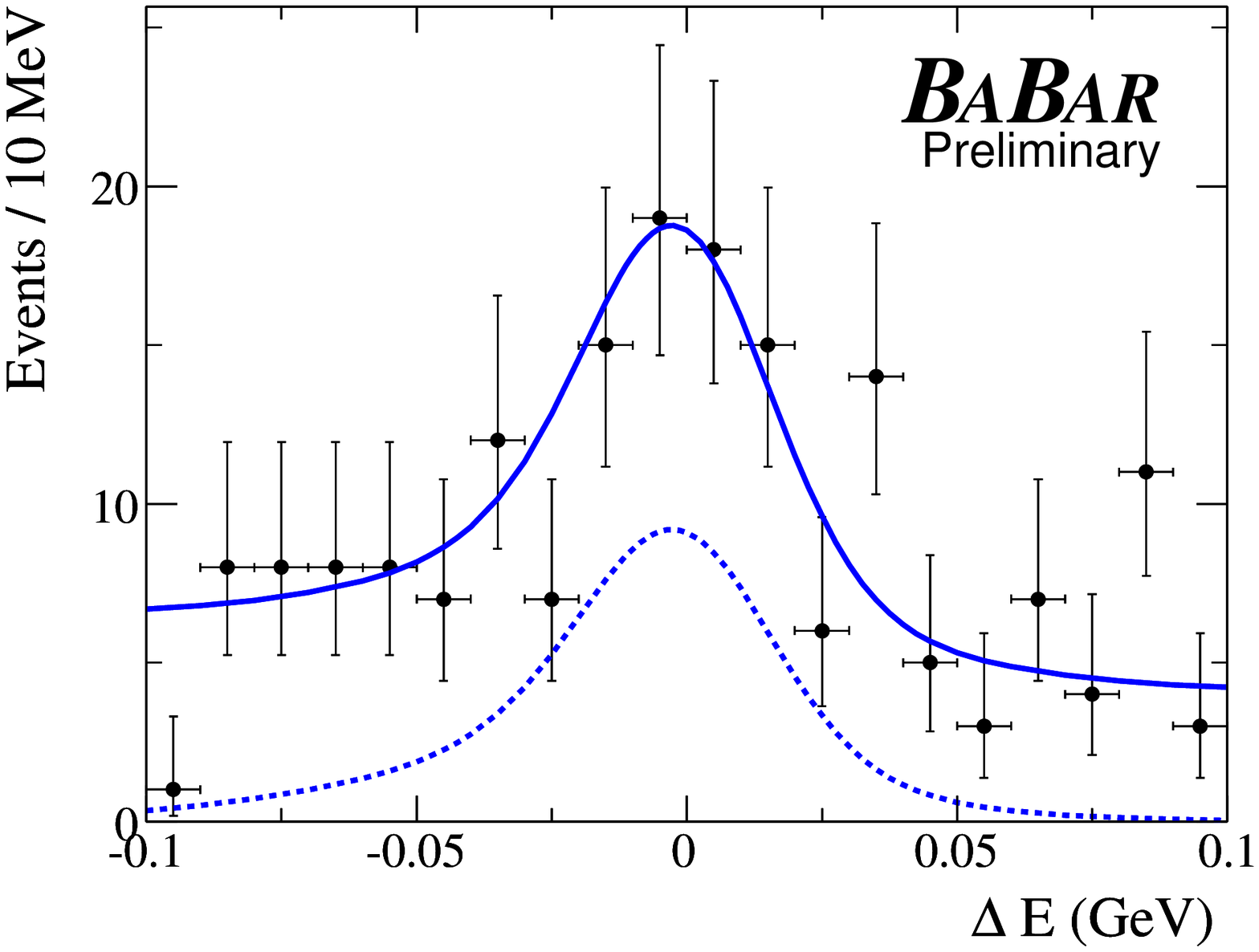}
\vspace{-.3cm}
 \caption{\label{Kstprhoz_proj}Projection plots of \mes\ (left) and \DE\
(right) for the $\Bp\to\rhoz\Kstarp$ (top) and $Bp\to f_0\Kstarp$ (bottom) 
samples.  The dotted lines show the signal fit component while the
solid lines are the full fit projection including signal and background.}
\end{figure}

\babar\ has a new measurement for this channel.  Both the $\Kstarp\to K\piz$
and $\Kstarp\to\KS\pip$ channels are used.  In this case there are
complications for both the $K\pi$ and $\pip\pim$ mass distributions.
The former has the same $K\pi$ S-wave contributions as the previous
analysis, while for $\pip\pim$, there are contributions for
$f_0(980)$ and $f_0(1370)$.  In Fig.~\ref{Kstprhoz_proj}, we show
projection plots for the $Bp\to\rhoz\Kstarp$ and $Bp\to f_0\Kstarp$
signals.  The decay $Bp\to f_0\Kstarp$ is observed for the first time with 
about 40 signal events in each \Kstarp\ decay channel and a fit significance of
5.0$\sigma$ including systematic uncertainties.  The measured branching
fraction is $(5.2\pm1.2\pm0.6)\timesix$ and $\acp=-0.34\pm0.21$.  The
significance for the decay $\Bp\to\rhoz\Kstarp$ is only 2.6$\sigma$.
The branching fraction is $(3.6\pm1.7\pm0.8)\timesix$ leading to a 90\%
C.L. upper limit of $5.9\timesix$.  The value of \fL\ determined
by the fit is $\fL=0.91^{+0.22}_{-0.20}$ though this is not considered a
measurement for this decay since the signal itself is not significant.
Belle has not yet reported measurements for these decays.

\section{\boldmath{$\Bz\to\aone^+\rhom$}}

Since the decay $\Bz\to a_1^+\pim$ has been observed with a branching
fraction of about $40\timesix$ \cite{BABARa1pi,Bellea1pi} (see
contribution by Christos Touramanis to this conference), it seems
likely that the branching fraction for the $B\to a_1\rho$ decays might
also be large.  \babar\ has recently submitted for publication a search 
for the decay $B\to a_1^+\rhom$ \cite{a1rho}.  They find no significant 
signal and measure a 90\% C.L. upper limit of $61\timesix$.

\section{\boldmath{\omegaKp, \omegaKs, and \omegapip}}

An updated analysis of the decays \omegaKp, \omegaKs, and \omegapip\ was
recently submitted for publication by \babar\ \cite{omegaKs}.  The
results for the time-dependent \CP\ asymmetry have been
presented by Matt Graham at this conference.  The branching fractions,
significance, and charge asymmetries are given in Table \ref{tab:omegaK_results}.
These measurements supersede previous \babar\ measurements and are in
good agreement with the most recent results from Belle \cite{BelleomegaK}.
The results are also in reasonable agreement with theoretical
expectations.

\begin{table}[!htb]
\caption{\label{tab:omegaK_results}
Measured branching fraction \calB, significance $S$ (with systematic 
uncertainties included), and charge asymmetry \acp\ for the decays
\omegaKp, \omegaKs, and \omegapip.
}
\begin{tabular}{lccccc}
\dbline
Mode   & $\calB\ (10^{-6})$   & $S~(\sigma$) & \acp \\
\dbline
\fomegapip      & \romegapip  &  \somegapip  & \Aomegapip \\
\fomegaKp       & \romegaKp   &  \somegaKp   & \AomegaKp \\
\fomegaKz       & \romegaKz   &  \somegaKz   & --- \\
\dbline
\end{tabular}
\vspace{5mm}
\end{table}

\section{Conclusions}
We have reported many new measurements, mostly for rare charmless
$B$-meson decays. Particularly noteworthy are the first observation of
the decay $B\to\etapr\Kstar$ with the resulting constraints on possible 
singlet diagrams and $1/m_b$ terms in QCD factorization and the many 
new results for $B$ decays to pairs of vector mesons.  The latter are
helping in the understanding of the small value of \fL\ for
$B\to\phi\Kstar$ and are helping to reduce the penguin uncertainties for 
the measurement of $\alpha$ in $\Bz\to\rhop\rhom$ decays.

\bigskip % extra skip inserted
\begin{acknowledgments}
We thank Chris Hearty and the rest of the organizing committee for a
very interesting and enjoyable conference.  We also are grateful for
helpful discussions with Michael Gronau, Harry Lipkin, Jon Rosner, and
Jure Zupan.
\end{acknowledgments}

\bigskip % extra skip inserted

\end{document}